\documentclass[11pt]{article} 
\usepackage[a4paper, margin=1.25in]{geometry} 
\usepackage[T1]{fontenc}

\usepackage[usenames,dvipsnames,svgnames,x11names]{xcolor}
\usepackage[nocompress]{cite}

\usepackage[pdftex]{graphicx}
\DeclareGraphicsExtensions{.pdf}

\usepackage[cmex10]{amsmath}
\usepackage{amssymb}
\usepackage{mathtools}
\usepackage{amsfonts}

\usepackage{multirow}
\usepackage{booktabs}
\usepackage{colortbl}
\usepackage{rotating}

\usepackage{array}
\newcolumntype{L}[1]{>{\raggedright\let\newline\\\arraybackslash\hspace{0pt}}m{#1}}
\newcolumntype{C}[1]{>{\centering\let\newline\\\arraybackslash\hspace{0pt}}m{#1}}
\newcolumntype{R}[1]{>{\raggedleft\let\newline\\\arraybackslash\hspace{0pt}}m{#1}}

\newcommand{\Nconf}{N_{95}}
\newcommand{\Qservice}{Q_{\rho,\sigma}}

\usepackage[pdftex,colorlinks=true,linkcolor=blue,urlcolor=blue,citecolor=blue]{hyperref}

\usepackage{enumitem}

\begin{document}

\title{\huge SemQBench: Benchmarking Decomposition-Aware Runtime Policies for Quantum-HPC Search\thanks{To Appear in the Proceedings of the \textit{International Workshop for Software Frameworks and Workload Management on Quantum and High-Performance Computing (HPC) Ecosystems (SFWM)}, Co-located with IEEE/ACM Supercomputing (SC) Conference, November, 2026.}}

\author{
Chandrachud Pati and Yogesh Simmhan
\\~\\
\textit{Indian Institute of Science, Bangalore 560012 India}
\\~\\
\texttt{Email: \{chandrachudp, simmhan\}@iisc.ac.in}
}
\date{}

\maketitle

\begin{abstract}
Quantum processors are increasingly used as accelerators inside high-performance computing (HPC) workflows. A workflow runtime decides which quantum circuits to run, how to group them into provider jobs, and what to repeat after a failed attempt. Many search problems split into independent parts, but the runtime can use that structure only when the workflow exposes the parts and a validity check for each one. SemQBench measures what a runtime gains from this visibility. It compares seven execution policies for the same search problem, including two controls without amplitude amplification. The policies change one runtime choice at a time: whether to amplify, whether partial results may be combined across runs, whether the parts share one circuit, and what a retry keeps. We run them on the 20-qubit IQM Garnet processor using 106 distinct circuits in three jobs, and report circuit executions, active device time, and provider jobs separately before combining them in a cost model. Amplifying the problem as one whole circuit shows no statistically clear gain over random sampling and uses $6.7$--$104\times$ more active device time. Exposing the parts and collecting valid partial results across a fixed batch of runs cuts the executions needed for 95\% completion confidence by $2.20$--$4.00\times$, and this policy is never worse than the other amplified policies in executions, device time, or jobs; it is selected in 99 of 108 overhead settings. Fewer executions do not always mean less device time: on one workload, executions fall from thirteen to four while active device time rises from $7.44$ to $9.81~\mu$s. Exposing problem structure therefore gives the runtime a strong default policy and the measurements needed to decide when the simpler unamplified option is cheaper.
\end{abstract}

\section{Introduction}

Quantum-HPC applications are increasingly composed as workflows that coordinate classical tasks, quantum-circuit execution, provider jobs, and recovery after incomplete work~\cite{humble2021qchpc,beck2024integrating,mantha2024pilot,chundury2025qfw,miniskar2026qiris}. A workflow runtime decides what circuits to submit, how to group executions into provider jobs, and what work to repeat after an incomplete attempt. Since the quantum device only executes the circuits it receives and returns samples, the workflow interface determines which problem structure the runtime can use.

A Q-HPC workflow may include a \textit{quantum search kernel}, such as a Grover-style search over assignments to a Boolean formula~\cite{grover1996,boyer1998}. Some formulas decompose into independent parts. E.g., if two variable groups do not appear together in any clause, each group can be checked by a separate validator, similar to the connected-component structure used in satisfiability problems~\cite{biere2006components}.
Consider the five-variable formula
\begin{equation} x_1 \wedge \neg x_2 \wedge (x_1 \vee x_2) \wedge x_3 \wedge (x_4 \leftrightarrow x_5) \label{eq:example} \end{equation}
No conjunct contains variables from both $\{x_1,x_2\}$ and $\{x_3,x_4,x_5\}$. The formula therefore has two independent variable groups: the first has one satisfying assignment out of four, and the second has two out of eight. We call each group a \emph{component}. A component's \emph{validator} accepts exactly the measured assignments that satisfy that component's conjuncts. A complete solution requires one valid component assignment for every component.

The workflow can submit this search to the runtime in two forms. In the \emph{whole form}, it compiles the complete formula into one circuit and supplies one validator for a complete assignment. The runtime then sees only whether each full circuit execution solved the formula. In the \emph{component-visible form}, the workflow supplies the components and their validators. The runtime can then keep a valid assignment for one component even when the same execution misses another component, and it can combine valid component assignments found in different executions.

This interface choice changes how amplitude amplification and recovery are used. \textit{Amplitude amplification} increases the probability of measuring assignments accepted by an oracle or validator, but its effect depends on how many assignments are accepted and on the number of amplification rounds~\cite{grover1996,boyer1998}. With a whole-form circuit, amplification targets the complete formula and retries repeat the complete circuit. With decomposed components, amplification rounds can be selected per component since each can have a different number of valid assignments. The runtime can also decide whether to accept all component results in one execution, accumulate valid component assignments across executions, place components in one circuit or in separate circuits, and repeat only the components that remain unsolved. A direct comparison between whole-form and component-visible execution changes several of these choices at once, so a resource difference cannot be attributed to one of them.

Our proposed \textit{SemQBench} turns this direct comparison into seven controlled policies, including two round-zero controls. Each policy fixes circuit form, amplification, result collection, packaging, and retry behavior. The controls sample uniformly, while the five amplified policies change one runtime choice at a time. All policies come from the same source predicate, and semantic audits verify that the two interfaces accept the same complete solutions.

The policy definitions offer two findings. For the same bundled circuit, collecting valid component results across a batch cannot do worse than requiring every component to succeed in one execution. Useful amplification rounds also depend on each component's valid-assignment fraction. Hardware measurements are needed to determine whether these gains persist under noise and whether fewer executions also reduce device time or provider jobs.

We evaluate four exhaustively checkable workload--mapping conditions using three IQM Garnet jobs with 106 distinct circuits. Whole-form amplification does not show a clear gain over exact random sampling and uses $6.7$--$104\times$ more active device time than its measured controls. With visible components, bundled fixed aggregation reduces the point-estimate execution count by $2.20$--$4.00\times$ relative to exact round-zero aggregation. It is no worse than the other amplified policies on the three reported resources, and the overhead sweep selects it in 99 of 108 settings. The unequal workload shows why ``fewer'' need not mean ``cheaper'': executions fall from thirteen to four while active device time rises from $7.44$ to $9.81~\mu$s. These small workloads are designed to isolate policy effects and do not evaluate quantum advantage.

We make three contributions.
\begin{enumerate}[leftmargin=*]
\item We build whole-form and component-visible round-zero controls from exact valid-assignment fractions and measured hardware distributions (\S~\ref{sec:model}, \S~\ref{sec:results}).
\item We define seven policies, including two controls, that change one runtime choice at a time. The hardware results identify bundled fixed aggregation as the strongest tested amplified policy and show the remaining trade-off against unamplified component sampling (\S~\ref{sec:policies}, \S~\ref{sec:results}).
\item We provide a runtime-facing descriptor, semantic checks, run manifests, and a public artifact. We report executions, active device time, and provider jobs separately before combining them into one modeled cost (\S~\ref{sec:hardware}, \S~\ref{sec:results}).
\end{enumerate}

\section{Related Work}

\subsection{Quantum-HPC workflow systems}

Quantum-HPC systems coordinate quantum devices, simulators, and classical resources through workflow, task, and runtime interfaces~\cite{mantha2024pilot,chundury2025qfw,miniskar2026qiris,shehata2026openqse}.
These systems manage submission, backend selection, task execution, and checkpointing. SemQBench studies what search structure a workflow must expose for the runtime to select among the evaluated policies. A runtime given only a whole-form circuit can schedule and retry it, but cannot validate or retain partial component progress.

\subsection{Decomposition, circuit cutting, and search structure}

Connected-component decomposition is an established method for satisfiability problems (SAT)~\cite{biere2006components}. \textit{Distributed Grover methods} use the same search structure to build component-level quantum searches~\cite{qiu2022distributed,huang2026lowdepth}. SemQBench uses such decomposable predicates as controlled workloads, but it does not claim the decomposition rule as new. Its focus is the runtime consequence of exposing the components: whether valid component assignments must arrive in the same execution, can be accumulated across executions, can share one physical circuit, or can be retried selectively.

\textit{Circuit cutting} also decomposes quantum work, but it solves a different problem~\cite{peng2020cutting,tang2021cutqc,khare2023parallelizing,tejedor2025qdislib,pal2026noiseaware}.
Circuit cutting partitions an already constructed circuit and reconstructs an output distribution from fragment results. Prior studies also show that the resulting fragments create workflow and hardware-placement trade-offs in hybrid Q-HPC settings~\cite{khare2023parallelizing,tejedor2025qdislib,pal2026noiseaware}. SemQBench instead starts from independent formula components whose validators define separate valid assignments. No reconstruction of a cut circuit is required: a complete solution is obtained when every component has produced a valid assignment. As a result, the fixed-aggregation model in Eq.~\eqref{eq:fixed} applies to component-validator subsets, not to circuit-fragment reconstruction.

\subsection{Quantum benchmarks and round-zero controls}

Quantum Mini-Apps and \textit{application-oriented benchmarks} provide small, controlled workloads for evaluating quantum-HPC software and devices~\cite{saurabh2024miniapps,lubinski2023benchmarks}. Such workloads are useful because their scale and structure can be chosen to expose a specific system effect. SemQBench follows this approach but uses the same source predicate across all policies. This lets us compare whole-form and component-visible execution without changing the application-level problem.

The \textit{round-zero controls} provide the baseline for this comparison. Zero amplification rounds correspond to uniform sampling, and the exact valid-assignment fraction is known for the tested formulas, so SemQBench separates success available from random sampling from success attributable to amplification. This differs from comparisons that report only the final amplified success probability. $U$-$W$ controls whole-form execution, and $U$-$C$ controls component-visible aggregation.

\begin{figure}[t]
\centering
\includegraphics[width=0.98\textwidth]{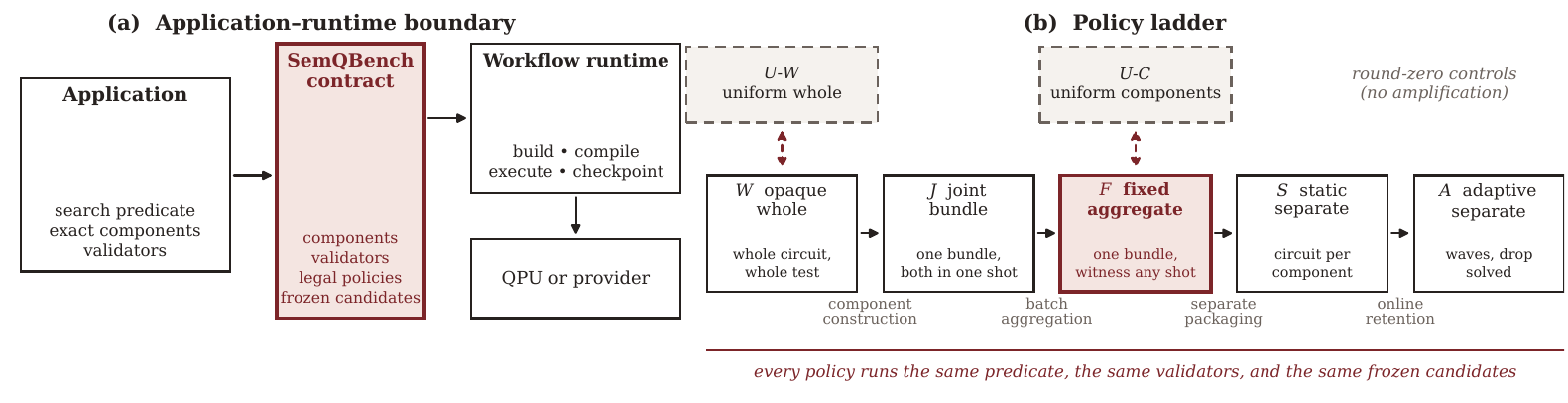}
\caption{SemQBench exposes components and validators to the workflow runtime (left) and compares seven policies, including two round-zero controls (right).}
\label{fig:stack}
\end{figure}

\subsection{Co-location and policy selection}

\textit{Quantum multiprogramming} co-locates circuits from unrelated programs and returns one result to each program~\cite{das2019multiprog,giortamis2025qos}.
The bundles in SemQBench have a different semantics. They contain components of one formula, and each component validator contributes to one complete search result. Thus, bundling is not a throughput optimization across independent users but rather a packaging choice for one decomposable workflow. Park et al. also execute decomposed Grover circuits concurrently, but optimize a single search probability rather than compare aggregation and retention policies that collect valid component assignments across executions~\cite{park2023grover}.

Reliability and selection models rank circuits, devices or placements using noise-aware compilation, fidelity estimation or learned predictors~\cite{murali2019noiseadaptive,quetschlich2024mqt,nguyen2025qfor}.
SemQBench is not a predictor. It measures a fixed set of policies and reports three resource measures: circuit executions, active device time, and provider jobs.
The distinction is relevant because a policy can reduce executions while increasing active device time. The overhead model is therefore applied only after these coordinates are reported separately.

\section{Application Boundary and Policies}\label{sec:policies}

Figure~\ref{fig:stack} shows the workflow boundary and the seven policies used in our evaluation. The two round-zero controls and five amplified policies compare amplification, aggregation, packaging, and retention.

\subsection{Application descriptor and run state}

The immutable descriptor records the formula, components, validators, legal policies, candidate round counts, and build settings. It is the runtime-facing input: it states what choices are allowed and freezes them before hardware outcomes are inspected. A semantic audit verifies that the whole-form and component validators accept the same complete solutions. A separate run manifest stores circuit identifiers, mappings, raw counts, calibration and job identifiers, checkpoints, and timings. Together, the two records associate each result with a fixed problem description and the corresponding hardware measurements.

\subsection{Workloads}

We choose small, exhaustively checkable workloads with known component structure so that resource differences can be attributed to the runtime policies being compared.

The \textit{equal-width $3+3$ workload} is a six-qubit exact-match search with one valid assignment and is evaluated on two hardware mappings. The other two workloads are Boolean formulas encoded in the DIMACS Conjunctive-Normal-Form (CNF) format~\cite{dimacs1993sat}. The \textit{unequal-width $2+3$ workload} has component satisfying-assignment counts $1$ and $2$, and two complete solutions. The \textit{three-component $2+2+2$ workload} has component satisfying-assignment counts $1$, $3$, and $2$, and six complete solutions. For each formula, the workload builder derives the connected components before circuit construction and emits both workflow forms. Hardware outcomes do not affect the component split.

\subsection{Policy definitions}

A \emph{shot} is one execution of a compiled circuit. A provider job may contain several circuits and many shots of each. For component $i$, let $p_i$ be the fraction of assignments accepted by its validator. With $r_i$ amplitude-amplification rounds, its noiseless success probability is $\sin^2((2r_i+1)\arcsin\sqrt{p_i})$ \cite{grover1996,boyer1998}. Each shot starts from the initial state. One \textit{round} maps valid-assignment fractions $1/4$, $1/2$, and $3/4$ to success probabilities $1$, $1/2$, and $0$, so the useful round count is component-specific and may be zero. \textit{Round zero} is uniform sampling.

The component-visible policies in Fig.~\ref{fig:stack} capture four runtime choices: how component results are validated, combined, packaged and retried. \emph{Joint-shot validation} requires every component validator to accept in the same shot. \emph{Fixed aggregation} may collect valid component assignments across a predetermined batch. A \emph{bundle} places all components in one physical circuit while retaining separate validators. Separated execution uses one component circuit at a time. A static policy executes predetermined work, whereas adaptive retention keeps completed components and re-executes only those still missing. Because hardware outcomes within a bundle need not be independent, bundled policies use the full measured distribution over validator subsets. Separated policies use independently measured component probabilities.

Combining these choices gives the seven policies used in the evaluation. $U$-$W$ and $W$ are the unamplified and amplified whole forms. $U$-$C$ is unamplified component-visible fixed aggregation. The amplified component-visible policies are $J$ (bundled joint-shot), $F$ (bundled fixed aggregation), $S$ (static separated), and $A$ (separated with adaptive retention).

\begin{table}[t]
\centering
\caption{Capabilities of the seven policies. Pairwise comparisons isolate amplification, aggregation, packaging, or retention.}
\label{tab:capabilities}
\small
\setlength{\tabcolsep}{4pt}
\begin{tabular}{lC{2.25cm}cC{1.6cm}cC{2.1cm}C{1.65cm}}
\toprule
\bf Policy & \bf Components visible & \bf Amplified & \bf Physical circuits & \bf Validation & \bf Fixed aggregation & \bf Adaptive retention\\
\midrule
$U$-$W$ & no & no & one & whole & no & no\\
$W$ & no & yes & one & whole & no & no\\
$U$-$C$ & yes & no & one & component & yes & no\\
$J$ & yes & yes & one & component & no & no\\
$F$ & yes & yes & one & component & yes & no\\
$S$ & yes & yes & multiple & component & yes & no\\
$A$ & yes & yes & multiple & component & yes & yes\\
\bottomrule
\end{tabular}
\end{table}

Table~\ref{tab:capabilities} summarizes the policy capabilities.
The pairwise comparisons isolate one runtime choice each. $U$-$W$/$W$ and $U$-$C$/$F$ isolate amplification, $J$/$F$ aggregation, $F$/$S$ packaging, and $S$/$A$ retention. For Eq.~\eqref{eq:example}, finding a valid assignment for the first component in one shot and for the second component in a later shot completes $F$ but not $J$. Mappings, round candidates, circuit forms, and component groupings are fixed before hardware execution. Policy $A$ replays measured component outcomes to evaluate retention and checkpointing; it does not use live adaptive resubmission.

\section{Probability and Resource Model}\label{sec:model}

SemQBench uses the same completion target for all policies: at least 95\% probability of recovering a complete solution. For whole-form policies, a complete solution must be produced by one circuit execution. For component-visible policies, different executions may produce valid assignments for different components. Thus, completion probability depends on how component results are combined, not only on the probability that one execution satisfies all validators.

\subsection{Round-zero controls}

At round zero, a width-$n$ circuit samples uniformly from its $2^n$ assignments. If $M$ assignments satisfy the complete predicate, the one-shot success probability is:
\begin{equation}
p_0=\frac{M}{2^n}. \label{eq:chance}
\end{equation}
This analytic value is reported separately from the measured $U$-$W$ control.

For $g$ independent components with exact valid-assignment fractions $p_i$, the ideal probability that exactly validator subset $B$ accepts in one bundled shot is:
\begin{equation}
q_B^{(0)}=\prod_{i\in B}p_i\prod_{i\notin B}(1-p_i). \label{eq:uniformsubset}
\end{equation}
The measured $U$-$C$ control instead uses the hardware histogram over all $2^g$ validator subsets. The analytic distribution is a calibration-independent reference and not bound by hardware behavior.

\subsection{Fixed aggregation}

Let $G=\{1,\ldots,g\}$ be the component set, and let $q_B$ be the measured probability that exactly the validators in $B\subseteq G$ accept in one bundled shot. After $k$ predetermined shots, fixed aggregation completes if every component has appeared in at least one accepting subset. By inclusion--exclusion, this probability is:
\begin{equation}
P_F(k)=\sum_{A\subseteq G}(-1)^{|A|}
\left(\sum_{B:B\cap A=\emptyset}q_B\right)^k.
\label{eq:fixed}
\end{equation}
The reported execution count is the smallest $k$ with $P_F(k)\geq0.95$. For two and three components, unit tests match Eq.~\eqref{eq:fixed} to direct sequence enumeration and verify monotonicity in $k$.

\subsection{Resource coordinates}

For one repeated circuit with valid-result probability $p$, the execution count needed for 95\% completion confidence is:
\begin{equation}
\Nconf(p)=\left\lceil\frac{\log(0.05)}{\log(1-p)}\right\rceil.
\label{eq:n95}
\end{equation}
For aggregation and retention policies, $\Nconf$ is the corresponding 0.95-quantile physical-circuit count. We report:
\begin{equation}
R_{95}=(\Nconf,D_{\mathrm{active}},H),
\label{eq:vector}
\end{equation}
where $D_{\mathrm{active}}$ is the sum of scheduled circuit durations and $H$ is the number of separately submitted provider jobs. This vector excludes reset delay and does not assign a monetary cost to provider jobs.

For overhead sensitivity only, we also use:
\begin{equation}
\Qservice=D_{\mathrm{active}}+\rho\Nconf+\sigma H.
\label{eq:q}
\end{equation}
The displayed reference is $Q_{250\mu\mathrm{s},0}$. The sweep uses $\rho\in\{0,50,100,250,500,1000\}~\mu$s per execution and $\sigma\in\{0,0.05,0.1,0.25,0.5,1,2,5,10\}$ ms per job, after unit conversion. This modeled quantity does not represent provider billing, queue delay, usage, or measured wall-clock time.

\subsection{Measurement sources and uncertainty}
Hardware probabilities, raw counts, compiled durations, and provider timelines are measured. Probabilities formed by multiplying separately measured component outcomes are derived from measurements, while exact chance values are analytic. Completion counts, $\Qservice$, and retention quantiles are modeled from the measured or analytic inputs.

Table~\ref{tab:resources} later reports exact binomial Clopper--Pearson intervals with a Bonferroni correction over 26 unique measured events~\cite{bonferroni1936,clopper1934}. The interval endpoints for $S$ and $A$ are products of component-level endpoints. The selected equal-width configurations also use a 98-event family containing every candidate examined. For the two CNF workloads, candidate choice uses one shot half and the reported probabilities use the disjoint held-out half. Controls use 1,024 shots, while amplified sweeps use 512 shots. The ceiling-quantized $\Nconf$ values are point-estimate summaries; The intervals support the direction of the reported differences but do not establish differences among the four reduction factors in Table~\ref{tab:chancecomparison}.

\section{Hardware Setup}\label{sec:hardware}

We execute all hardware circuits on \textit{IQM Garnet}, a 20-qubit square-lattice superconducting processor~\cite{abdurakhimov2024iqm,iqmqiskit}.
It provides enough qubits for all tested workloads while exposing the success probabilities, scheduled durations, and calibration effects that determine policy ordering.
Every job uses transpiler seed 42. Heralding, which flags some state-preparation errors, and dynamical decoupling, which inserts pulses during idle periods, are disabled. Thus, the measurements describe the plain compiled circuits.

The hardware evaluation uses three provider \textit{jobs}. The \textit{equal-width workload} job runs 44 distinct circuits with 512 shots on each of two hardware mappings. The \textit{extended} job runs 54 circuits with 512 shots across the unequal-width and three-component workloads. The \textit{round-zero} job adds eight circuits with 1,024 shots each. Together, these jobs cover 106 distinct circuits, with provider timelines of
9.33, 9.13 and 2.68
QPU-seconds. These timelines include all shots in each provider job, whereas the microsecond values reported later give the scheduled duration of one circuit execution.

The unequal-width and three-component controls share a calibration set with their amplified runs. For the equal-width workload, the measured $U$-$C$ controls and amplified candidates come from different hardware records, so their measured comparison is descriptive. The exact analytic $U$-$C$ control is calibration-independent, but it is not a hardware measurement.

The extended job repeats reference circuits, or anchors, before and after its sweep. Endpoint probability changes are at most $0.031$ for component circuits and 0.020 for bundled circuits. The two whole-form endpoint changes are $-0.020$ and $+0.039$. These anchor endpoints provide a local calibration check. They do not bound unobserved fluctuations between the two anchor executions, and they do not describe another day or device.

We use two compiler constructions for the whole predicate. The first uses Qiskit's \texttt{PhaseOracleGate}, and the second uses a truth-table diagonal phase~\cite{qiskit,qiskitphaseoracle,qiskitdiagonal,qiskitgrover}. Statevector checks show that the two constructions agree on marked assignments and probabilities to $1.68\times10^{-14}$.
A reproducible artifact with the descriptors, manifests, raw counts, mappings, tests, and regeneration scripts is available at \url{https://github.com/dream-lab/semqbench}.

\section{Results}\label{sec:results}

We evaluate SemQBench in the order of the policy comparisons in Fig.~\ref{fig:stack}. We first summarize the full resource table for all policies. We then compare whole-form amplification with its round-zero control, followed by component-visible amplification, aggregation, packaging, and retention. The final subsections examine the overhead model and the whole-form compiler construction.

\subsection{Primary resource results}

\begin{figure}[t]
\centering
\includegraphics[width=\textwidth]{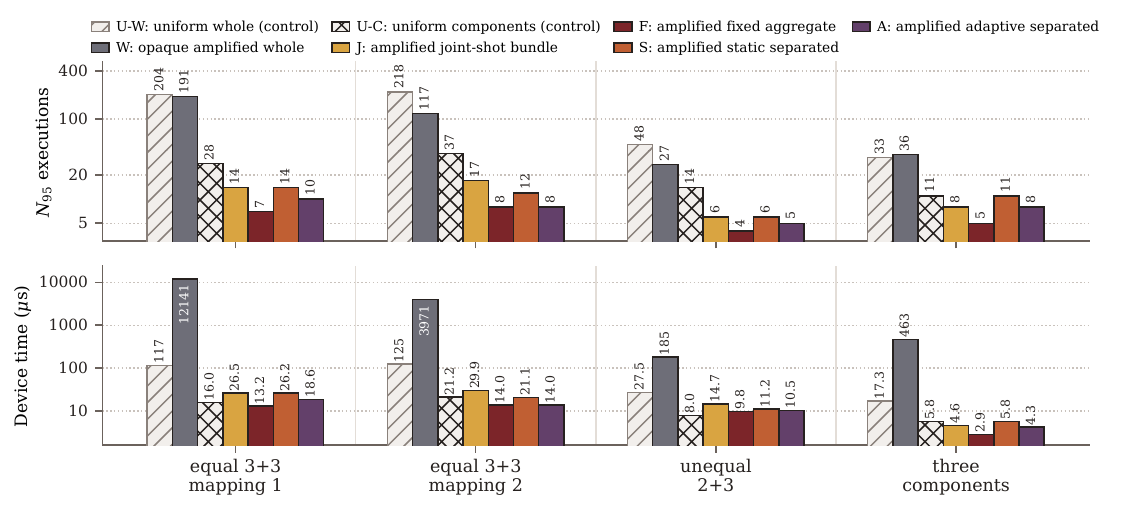}
\caption{Executions and active device time required for 95\% completion confidence. Hatched bars denote round-zero controls.}
\label{fig:resources}
\end{figure}

Figure~\ref{fig:resources} and Table~\ref{tab:resources} report the seven policies across four workload--mapping conditions. The $p_0$ column gives the exact complete-assignment probability under uniform sampling. For $U$-$W$, $W$, $U$-$C$, $J$, and $F$, the measured probability column is estimated from one physical circuit. For $S$ and $A$, no single circuit contains every component, so the comparable one-attempt probability is the product of separately measured component probabilities and is marked with a $\dagger$.

The completion count depends on how each policy uses circuit outcomes. For $J$, the joint-valid probability enters Eq.~\eqref{eq:n95}. For $U$-$C$ and $F$, $\Nconf$ is the fixed-batch count computed from the measured validator-subset distribution using Eq.~\eqref{eq:fixed}. Policy $S$ applies fixed aggregation to separately measured component circuits, and $A$ uses the adaptive-retention replay. The column $\bar d=D_{\mathrm{active}}/\Nconf$ reports the mean scheduled duration per counted physical-circuit execution. The valid/$n$ entries give the raw successes and shot counts.

\begin{sidewaystable}
\centering
\caption{Measured probabilities and resource coordinates for the seven policies. Dagger ($\dagger$) marks probabilities that are products of separately measured component probabilities.}
\label{tab:resources}
\footnotesize
\setlength{\tabcolsep}{4pt}
\begin{tabular}{L{2.1cm}lrccR{2.0cm}rrrrR{1.7cm}}
\toprule
\bf Condition & \bf Policy & $p_{0}$ & $\widehat p_{\rm valid}^{(1)}~[L,U]$ & \bf Valid outcomes / shots & \bf Success relative to chance & $N_{95}$ & $\bar d$ ($\mu$s) & $D_{\rm active}$ ($\mu$s) & $H$ & \bf Modeled $Q_{250,0}$ (ms)\\
\midrule
\bf Equal 3+3, mapping 1 & U-W & 0.0156 & 0.0146 $[0.0057,0.0303]$ & $15/1024$ & 0.94 & 204 & 0.572 & 116.69 & 1 & 51.117\\
 & W & 0.0156 & 0.0156 $[0.0038,0.0409]$ & $8/512$ & 1.00 & 191 & 63.564 & 12140.72 & 1 & 59.891\\
 & U-C & 0.0156 & 0.0176 $[0.0075,0.0343]$ & $18/1024$ & 1.12 & 28 & 0.572 & 16.02 & 1 & 7.016\\
 & J & 0.0156 & 0.1934 $[0.1426,0.2524]$ & $99/512$ & 12.38 & 14 & 1.892 & 26.49 & 1 & 3.526\\
 & F & 0.0156 & 0.1934 $[0.1426,0.2524]$ & $99/512$ & 12.38 & 7 & 1.892 & 13.24 & 1 & 1.763\\
 & S & 0.0156 & 0.1841$^\dagger$ $[0.1307,0.2485]$ & $(245\cdot197)/(512^2)$ & 11.78 & 14 & 1.874 & 26.23 & 1 & 3.526\\
 & A & 0.0156 & 0.1841$^\dagger$ $[0.1307,0.2485]$ & $(245\cdot197)/(512^2)$ & 11.78 & 10 & 1.863 & 18.63 & 9 & 2.519\\ \midrule
\addlinespace[2pt]
\bf Equal 3+3, mapping 2 & U-W & 0.0156 & 0.0137 $[0.0051,0.0290]$ & $14/1024$ & 0.88 & 218 & 0.572 & 124.70 & 1 & 54.625\\
 & W & 0.0156 & 0.0254 $[0.0090,0.0549]$ & $13/512$ & 1.62 & 117 & 33.940 & 3970.98 & 1 & 33.221\\
 & U-C & 0.0156 & 0.0117 $[0.0039,0.0263]$ & $12/1024$ & 0.75 & 37 & 0.572 & 21.16 & 1 & 9.271\\
 & J & 0.0156 & 0.1641 $[0.1170,0.2202]$ & $84/512$ & 10.50 & 17 & 1.756 & 29.85 & 1 & 4.280\\
 & F & 0.0156 & 0.1641 $[0.1170,0.2202]$ & $84/512$ & 10.50 & 8 & 1.756 & 14.05 & 1 & 2.014\\
 & S & 0.0156 & 0.2362$^\dagger$ $[0.1739,0.3083]$ & $(231\cdot268)/(512^2)$ & 15.11 & 12 & 1.756 & 21.07 & 1 & 3.021\\
 & A & 0.0156 & 0.2362$^\dagger$ $[0.1739,0.3083]$ & $(231\cdot268)/(512^2)$ & 15.11 & 8 & 1.756 & 14.05 & 6 & 2.014\\ \midrule
\addlinespace[2pt]
\bf Unequal 2+3 & U-W & 0.0625 & 0.0605 $[0.0399,0.0872]$ & $62/1024$ & 0.97 & 48 & 0.572 & 27.46 & 1 & 12.027\\
 & W & 0.0625 & 0.1051 $[0.0546,0.1768]$ & $27/257$ & 1.68 & 27 & 6.836 & 184.57 & 1 & 6.935\\
 & U-C & 0.0625 & 0.0615 $[0.0407,0.0883]$ & $63/1024$ & 0.98 & 14 & 0.572 & 8.01 & 1 & 3.508\\
 & J & 0.0625 & 0.4032 $[0.3092,0.5024]$ & $102/253$ & 6.45 & 6 & 2.452 & 14.71 & 1 & 1.515\\
 & F & 0.0625 & 0.4032 $[0.3092,0.5024]$ & $102/253$ & 6.45 & 4 & 2.452 & 9.81 & 1 & 1.010\\
 & S & 0.0625 & 0.5823$^\dagger$ $[0.4613,0.6913]$ & $(231\cdot173)/(246\cdot279)$ & 9.32 & 6 & 1.863 & 11.18 & 1 & 1.511\\
 & A & 0.0625 & 0.5823$^\dagger$ $[0.4613,0.6913]$ & $(231\cdot173)/(246\cdot279)$ & 9.32 & 5 & 2.098 & 10.49 & 4 & 1.260\\ \midrule
\addlinespace[2pt]
\bf Three components & U-W & 0.0938 & 0.0869 $[0.0619,0.1175]$ & $89/1024$ & 0.93 & 33 & 0.524 & 17.29 & 1 & 8.267\\
 & W & 0.0938 & 0.0810 $[0.0369,0.1487]$ & $20/247$ & 0.86 & 36 & 12.860 & 462.96 & 1 & 9.463\\
 & U-C & 0.0938 & 0.0889 $[0.0636,0.1197]$ & $91/1024$ & 0.95 & 11 & 0.524 & 5.76 & 1 & 2.756\\
 & J & 0.0938 & 0.3415 $[0.2511,0.4407]$ & $84/246$ & 3.64 & 8 & 0.572 & 4.58 & 1 & 2.005\\
 & F & 0.0938 & 0.3415 $[0.2511,0.4407]$ & $84/246$ & 3.64 & 5 & 0.572 & 2.86 & 1 & 1.253\\
 & S & 0.0938 & 0.3374$^\dagger$ $[0.2173,0.4732]$ & $(216\cdot184\cdot130)/(246\cdot247\cdot252)$ & 3.60 & 11 & 0.524 & 5.76 & 1 & 2.756\\
 & A & 0.0938 & 0.3310$^\dagger$ $[0.2118,0.4667]$ & $(216\cdot184\cdot125)/(246\cdot247\cdot247)$ & 3.53 & 8 & 0.535 & 4.28 & 4 & 2.004\\
\bottomrule
\end{tabular}
\end{sidewaystable}

Repeated duration entries in Table~\ref{tab:resources} come from independently compiled schedules rather than copied values. For the three-component workload, the $S$ policy's weighted mean over component durations rounds to the same value that happens to equal the $0.524~\mu$s round-zero bundle.

\subsection{Whole-form amplification versus uniform sampling}\label{sec:wholeamp}

The first comparison is $U$-$W$/$W$. Both policies use the whole-form interface, and only $W$ applies amplitude amplification. We treat an amplified configuration as \textit{resolved above a reference} only if its corrected confidence interval lies entirely above that reference.

For the whole-form interface, none of the four selected amplified configurations is resolved above exact random sampling. For the equal-width workload, the selection-aware 98-event intervals are $[0.0031,0.0447]$ on mapping 1 and $[0.0078,0.0591]$ on mapping 2; both overlap the exact chance value $0.0156$. For the unequal-width and three-component workloads, the disjoint evaluation intervals are $[0.0546,0.1768]$ and $[0.0369,0.1487]$; these overlap the chance values $0.0625$ and $0.0938$. At the point estimate, equal-width mapping 1 matches chance, and the three-component configuration is below chance.

The measured $U$-$W$ controls are also below exact chance in all four conditions, at $0.94$, $0.88$, $0.97$, and $0.93$ of the analytic values. Relative to these measured controls, the $W$ point estimates are $1.07$, $1.85$, and $1.74\times$ larger on the two equal-width mappings and the unequal-width workload, and $0.93\times$ the control on the three-component workload. The corrected $W$ and $U$-$W$ intervals overlap in every condition. Thus, these measurements do not resolve an improvement over uniform whole-form sampling. They also do not establish equivalence to chance or prove that amplification has no effect.

The resource measurements show a larger difference because the amplified whole-form circuits have greater compiled depth. Policy $W$ lowers the point-estimate $\Nconf$ on equal-width mapping 2 and the unequal-width workload, but active device time rises from $124.70$ to $3970.98~\mu$s and from $27.46$ to $184.57~\mu$s. On the three-component workload, $W$ requires 36 executions and $462.96~\mu$s, compared with 33 executions and $17.29~\mu$s for $U$-$W$. Across all four conditions, $W$ uses $6.7$--$104.0\times$ the active device time of the measured control without a resolved probability improvement. This observation applies to the two whole-predicate compiler constructions evaluated in Sec.~\ref{sec:wholeconstruction}; it may not apply to other whole-form constructions.

The held-out split also changes the reported point estimate. For the three-component $W$ circuit, the selection half gives $0.136$ and the evaluation half gives $0.081$. Reporting the selection half would therefore raise the point estimate by approximately 68\% relative to the held-out value.

\subsection{Component-visible amplification}

The component-visible amplification comparison compares $U$-$C$ with $F$. Both policies expose component validators, place the components in one bundled circuit, collect valid component assignments across a predetermined batch, and use one provider job. They differ only in whether amplitude amplification is applied. For the unequal-width and three-component workloads, the two policies share a calibration record. For the equal-width mappings, the measured records differ, so the measured comparison is descriptive; the exact round-zero distribution provides the calibration-independent reference.

The selected round tuples are $(1,1)$ for both equal-width mappings, $(1,1)$ for the unequal-width workload, and $(1,0,0)$ for the three-component workload. Candidate sets are frozen before execution. The selected tuple minimizes $Q_{250\mu\mathrm{s},0}$ on the selection shots, with ties broken by tuple order. Thus selection uses $\rho=250~\mu$s per execution and $\sigma=0$ per job. For the two CNF workloads, selection and evaluation use disjoint shot halves. Table~\ref{tab:idealreference} compares the exact round-zero, ideal amplified, and measured amplified probabilities. Statevector execution of the audited component oracles independently reproduces the ideal values from the standard Grover expression.

\begin{table}[t]
\centering
\caption{Selected bundled fixed-aggregation circuit.}
\label{tab:idealreference}
\small
\begin{tabular}{lccccc}
\toprule
\bf Condition & \bf Rounds & $p_0$ & $p_{\rm ideal}$ & $\widehat p$ & $\widehat p/p_{\rm ideal}$\\
\midrule
\bf Equal $3+3$, mapping 1 & $(1,1)$ & 0.0156 & 0.6104 & 0.1934 & 0.317\\
\bf Equal $3+3$, mapping 2 & $(1,1)$ & 0.0156 & 0.6104 & 0.1641 & 0.269\\
\bf Unequal $2+3$ & $(1,1)$ & 0.0625 & 1.0000 & 0.4032 & 0.403\\
\bf Three components & $(1,0,0)$ & 0.0938 & 0.3750 & 0.3415 & 0.911\\
\bottomrule
\end{tabular}
\end{table}

The three-component workload illustrates the effect of selecting round counts separately for each component. Its component valid-assignment fractions are $\frac{1}{4}$, $\frac{3}{4}$ and $\frac{1}{2}$. One amplification round improves only the first component, so the selected tuple is $(1,0,0)$. Applying one round to all three components would reduce the ideal joint-valid probability from $0.3750$ to zero.

\begin{table}[t]
\centering
\caption{Bundled fixed aggregation $F$ compared with component-visible round-zero aggregation $U$-$C$.}
\label{tab:chancecomparison}
\small
\setlength{\tabcolsep}{4pt}
\begin{tabular}{L{2.3cm}R{1.25cm}R{1.8cm}R{0.8cm}R{1.5cm}R{1.8cm}R{1.5cm}R{1.2cm}}
\toprule
\bf Condition & \bf Exact $U$-$C$ $N_{95}$ & \bf Measured $U$-$C$ $N_{95}$ & \bf $F$ $N_{95}$ & \bf Count Reduc. vs. Exact & \bf Count Reduc. vs. Measured & \bf Exact $U$-$C$ active ($\mu$s) & \bf $F$ active ($\mu$s)\\
\midrule
\bf Equal $3+3$, mapping 1 & 28 & 28 & 7 & 4.00$\times$ & 4.00$\times$ & 16.02 & 13.24\\ \midrule
\bf Equal $3+3$, mapping 2 & 28 & 37 & 8 & 3.50$\times$ & 4.63$\times$ & 16.02 & 14.05\\ \midrule
\bf Unequal $2+3$ & 13 & 14 & 4 & 3.25$\times$ & 3.50$\times$ & 7.44 & 9.81\\ \midrule
\bf Three components & 11 & 11 & 5 & 2.20$\times$ & 2.20$\times$ & 5.76 & 2.86\\
\bottomrule
\end{tabular}
\end{table}

Table~\ref{tab:chancecomparison} reports the calibration-independent count comparison first. Relative to exact $U$-$C$, policy $F$ reduces the point-estimate $\Nconf$ by $2.20$--$4.00\times$, from $28$ to $7$, $28$ to $8$, $13$ to $4$, and $11$ to $5$. Relative to measured $U$-$C$, the descriptive range is $2.20$--$4.63\times$. The $4.63\times$ value occurs on equal-width mapping 2, where the measured control and amplified candidates use different hardware records, so it is not a calibration-matched effect. These factors refer only to the circuit-execution count.

The probability results show the same trend, although the calibration distinction limits the comparisons that can be made directly. The measured $U$-$C$ and $F$ intervals are disjoint in all four conditions, but only the unequal-width and three-component comparisons share a calibration record. For the equal-width mappings, the measured comparison remains descriptive. Against the exact analytic control, every $F$ interval lies above round-zero probability. The two equal-width bundles were selected from a sweep, and both remain above chance under the wider selection-aware family defined in Sec.~\ref{sec:model}.

The lower execution count does not always reduce active device time. Relative to exact $U$-$C$, $F$ lowers active time on both equal-width mappings and on the three-component workload. On the unequal-width workload, however, the count falls from thirteen to four while active time rises from $7.44$ to $9.81~\mu$s. Against measured $U$-$C$, the same case is fourteen executions and $8.01~\mu$s versus four executions and $9.81~\mu$s. Neither policy dominates the other in $(\Nconf,D_{\mathrm{active}},H)$ for that condition.

The anchor records provide a local check on the two same-calibration comparisons. In the extended job, the largest component-anchor endpoint change is $0.031$, while the amplified-minus-round-zero one-shot gaps are $0.342$ for the unequal-width workload and $0.253$ for the three-component workload. Both gaps also exceed $0.039$, the largest endpoint change among all recorded anchors in that job. This comparison is descriptive and does not establish replication across calibration windows, days or devices.

\subsection{Joint-shot versus fixed aggregation}

To isolate aggregation, we compare $J$ with $F$. Both policies use the same selected bundled circuit and the same measured validator-subset distribution. They differ only in how component results are combined. Policy $J$ requires all component validators to accept in the same shot. Policy $F$ may collect valid component assignments across a fixed batch.

Let $P_J(k)$ be the probability that at least one of $k$ shots has every validator accept in the same shot. Let $P_F(k)$ be the fixed-aggregation probability from Eq.~\eqref{eq:fixed}. Every completion under $J$ is also a completion under $F$, because a shot that accepts all validators also supplies all component assignments for the batch. Thus, $P_F(k)\geq P_J(k)$ for every $k$, and $\Nconf$ cannot favor $J$ for the same bundled circuit.

The measurements quantify the reduction in execution count. Policy $F$ reduces $\Nconf$ from $14$ to $7$, $17$ to $8$, $6$ to $4$, and $8$ to $5$ across the four workload--mapping conditions. Because $J$ and $F$ use the same circuit duration and provider-job count within each condition, active device time falls in the same proportions.

\subsection{Bundled versus separated packaging}

To isolate packaging, we compare $F$ with $S$. Both policies expose component validators and use fixed aggregation. However, policy $F$ places all components in one bundled circuit, while $S$ executes one component circuit at a time. Policy $F$ uses fewer physical-circuit executions on both equal-width mappings and on the unequal-width workload.

Execution counting differs between the bundled and separated policies for the three-component workload. Policy $S$ uses eleven component-circuit executions. Policy $F$ repeats a three-component bundle five times, so those five physical executions contain fifteen component instances. The reported comparison is therefore five bundled executions versus eleven separate component-circuit executions, not fifteen versus eleven. Bundling keeps component validation separate without requiring a separate physical execution for each component.

For $S$ and $A$, the model multiplies separately measured component probabilities and does not include cross-submission drift or provider failures. Even under that favorable assumption for separated execution, $F$ uses fewer executions than $S$ in all four conditions and fewer than $A$ in three, tying $A$ once. On equal-width mapping 2, $S$ has the larger one-attempt product, $0.2362$ versus $0.1641$, but still requires twelve physical-circuit executions rather than eight because each $F$ shot can contribute valid assignments for both components.

\subsection{Static versus adaptive retention}

To isolate retention, we compare $S$ with $A$. Both policies execute separate component circuits. Policy $S$ follows a fixed schedule, while $A$ keeps completed components and re-executes only the components that remain unsolved. Policy $A$ replays measured component outcomes to evaluate retention and checkpointing; it does not use live adaptive resubmission.

The replay lowers the 0.95-quantile execution count from $14$ to $10$, $12$ to $8$, $6$ to $5$, and $11$ to $8$. It also raises provider jobs from one to $9$, $6$, $4$, and $4$ in the corresponding modeled traces. Adaptive retention therefore reduces circuit count, but it does not dominate static execution when provider jobs carry overhead.

\subsection{Overhead sensitivity}

Equation~\eqref{eq:q} combines execution count, active device time, and provider jobs only after per-execution and per-job overheads are chosen. Across the 108 tested settings, bundled fixed aggregation $F$ is selected 99 times: 45 of 54 settings for the unequal-width workload and all 54 settings for the three-component workload. The unamplified component control $U$-$C$ is selected in the remaining nine settings, all with zero per-execution overhead. No other policy is selected.

\begin{table}[t]
\centering
\caption{Policies that minimize the modeled cost over the tested overhead grid.}
\label{tab:map}
\small
\begin{tabular}{lrrr}
\toprule
\bf Workload & \bf Cells & \bf $U$-$C$ selected & \bf $F$ selected\\
\midrule
\bf Three components & 54 & 0 & 54\\
\bf Unequal 2+3 & 54 & 9 & 45\\
\bottomrule
\end{tabular}
\end{table}

The policy sweep selects $F$ in 99 of the 108 tested overhead settings. Across all four conditions, $F$ is no worse than $U$-$W$, $W$, $J$, $S$, or $A$ on executions, active device time, and jobs, and it is better in at least one measure. It therefore performs at least as well as every other tested amplified policy on the three reported resource measures. The only remaining cost trade-off is $F$ against $U$-$C$ on the unequal-width workload. Both use one job, and their modeled costs cross at only $\rho=(9.81-8.01)/(14-4)=0.18~\mu\mathrm{s}$ per execution. With zero execution overhead, $U$-$C$ is cheaper; above that point, $F$ is cheaper in this model.

\subsection{Whole-form construction}\label{sec:wholeconstruction}

The second whole-predicate compiler construction checks whether the whole-form result is only an artifact of one lowering; attributing such losses to individual compilation stages is studied elsewhere~\cite{pati2026hbr}. For the unequal-width workload, the truth-table diagonal lowering reduces native depth from 356 to 221 and cross-component entangling gates from 56 to 52. For the three-component workload, depth falls from 965 to 431 and cross-component gates from 301 to 157. Across the full 512-shot candidate-sweep records, the measured valid-result point estimate rises from $0.064$ to $0.104$ and from $0.066$ to $0.109$. These records include the shots used for selection, unlike the disjoint evaluation halves in Table~\ref{tab:resources}, so they are not part of the interval test in Sec.~\ref{sec:wholeamp}.

The selected component bundles contain no cross-component two-qubit gates. In the same full candidate-sweep records, their representative round tuples have one-shot joint-valid probabilities $0.398$ and $0.314$. The held-out evaluation halves give $0.4032$ and $0.3415$, the values used in the main comparison. The alternative whole-form lowering reduces the full-record point-estimate gap, but a gap remains for both workloads.

\section{Discussion} \label{sec:discussion}

The main result concerns the relative performance of the tested amplified policies. When amplification is used, bundled fixed aggregation performs at least as well as the other tested amplified policies on executions, active device time, and provider jobs, and performs better on at least one measure. The remaining comparison is between amplified and unamplified component sampling. On the unequal-width workload, amplification saves ten executions but adds $1.80~\mu$s of active device time; hence, its modeled benefit depends on the per-execution overhead.

Two findings are structural rather than hardware discoveries: fixed aggregation cannot do worse than joint-shot validation for the same bundled circuit, and useful round counts depend on each component's valid fraction. The hardware measurements provide three additional results: whole-form amplification does not show a clear gain, compiled circuit cost can reverse a reduction in execution count, and additional provider jobs can offset the benefit of adaptive retry.

The SemQBench descriptor records the legal policies, while the manifest reports separate measured values for executions, device time, and jobs that can inform scheduler decisions. The current study evaluates fixed schedules rather than a live online scheduler. Its scope is also limited to exhaustively checkable workloads, zero or one amplification round, and three IQM Garnet jobs. The policy order may change for deeper circuits, more components, or another device.

\section{Conclusion}

A quantum-HPC runtime can use independent search components only when the workflow exposes the components and their validators. SemQBench studies this interface with seven policies, including two round-zero controls. Across four workload--mapping conditions, whole-form amplification does not show a clear gain over exact random sampling, while component-visible amplification reduces the point-estimate fixed-batch execution count by $2.20$--$4.00\times$ relative to exact round-zero aggregation. Bundled fixed aggregation is the strongest tested amplified policy and is selected in 99 of 108 modeled-cost settings. The remaining trade-off is against unamplified component sampling on one workload, where executions fall from thirteen to four but active device time rises from $7.44$ to $9.81~\mu$s. Exposing components therefore gives the runtime both a strong amplified default and the measurements needed to decide whether amplification is worth its added circuit time.

Future work will extend SemQBench from fixed-policy comparisons to adaptive scheduling of decomposable search within larger quantum-HPC workflows.
Using the exposed components and validators, we will study how runtimes can jointly choose amplification rounds, circuit bundles, and resubmission batches as search progress, device conditions, and provider costs change.
The broader goal is to coordinate classical validation and quantum execution, retaining validated components while reducing the total time and resources needed to obtain a complete valid solution.

\section*{Acknowledgment}
This work was supported by a grant towards the SPAARQ project from the National Quantum Mission (NQM), Department of Science \& Technology (DST), Government of India. 

Generative AI tools were used to improve the clarity of author-written text, and not for new technical content or data. All edits were reviewed and verified by the authors.


\bibliographystyle{plain}
\bibliography{references}

\end{document}